\documentclass[final,3p,times,onecolumn]{elsarticle}
\usepackage{lineno,hyperref}

\usepackage{amsmath}
\usepackage{graphicx,epsfig,wrapfig}
\usepackage[caption=false]{subfig}

\usepackage{amsthm,mathrsfs}
\usepackage{makeidx}
\usepackage{amsfonts}
\usepackage{amssymb}
\usepackage{amsmath}
\usepackage{mathtools}
\usepackage{fixmath}
\usepackage{amsbsy}
\usepackage[capitalize]{cleveref}
\usepackage{mathrsfs}
\usepackage{slashed}
\usepackage{bm}
\usepackage{multirow}
\usepackage{booktabs}
\usepackage{adjustbox}

\usepackage{color}
\usepackage[normalem]{ulem}

\usepackage[absolute]{textpos}
\textblockorigin{5mm}{5mm}

\definecolor{myGreen}{rgb}{0.2,0.72,0.2}
\renewcommand\sout{\bgroup \color[rgb]{0.55,0.00,0.99} \ULdepth=-.5ex \ULset}

\usepackage{simplewick}

\usepackage{multirow}
\usepackage{braket}

\newcommand{\polini}{\epsilon}
\newcommand{\polfin}{\epsilon'^*}
\newcommand{\polinimu}[1]{\epsilon^{#1}}
\newcommand{\polfinmu}[1]{\epsilon'^{*#1}}

\renewcommand{\[}{\begin{equation}}
\renewcommand{\]}{\end{equation}}

\definecolor{darkgreen}{RGB}{0,120,0}

\definecolor{orange}{RGB}{255,165,0}

\allowdisplaybreaks

\modulolinenumbers[5]
\journal{Physics Letters B}
\biboptions{sort&compress}
\begin{document}


\begin{textblock}{4}(16,0)
\begin{itemize}
\itemsep-0.5em
    \item[] DESY-24-117
    \item[] JLAB-THY-24-4130
\end{itemize}
\end{textblock}

\begin{frontmatter}

\title{Perturbative results of matrix elements of the axial current \\
and their relation with the axial anomaly}
\author[JCaddr]{Ignacio Castelli}
\ead{jorge.castelli@temple.edu}
\author[AFaddr]{Adam Freese}
\ead{afreese@jlab.org}
\author[CLaddr]{C\'edric Lorc\'e}
\ead{cedric.lorce@polytechnique.edu}
\author[JCaddr]{Andreas Metz}
\ead{metza@temple.edu}
\author[BPaddr1,BPaddr2]{Barbara Pasquini}
\ead{barbara.pasquini@unipv.it}
\author[SRaddr]{Simone Rodini}
\ead{simone.rodini@desy.de}

\address[JCaddr]{Department of Physics, SERC, Temple University, Philadelphia, PA 19122, USA}
\address[AFaddr]{Theory Center, Jefferson Lab, Newport News, Virginia 23606, USA}
\address[CLaddr]{CPHT, CNRS, \'Ecole polytechnique, Institut Polytechnique de Paris, 91120 Palaiseau, France}
\address[BPaddr1]{Dipartimento di Fisica, Universit\`a degli Studi di Pavia, 27100 Pavia, Italy}
\address[BPaddr2]{Istituto Nazionale di Fisica Nucleare, Sezione di Pavia, 27100 Pavia, Italy}
\address[SRaddr]{Deutsches Elektronen-Synchrotron DESY, Notkestr. 85, 22607 Hamburg, Germany}

\begin{abstract} 
In the Standard Model of particle physics, the axial current is not conserved,
due both to fermion masses and to the axial anomaly.
Using perturbative quantum chromodynamics, we calculate matrix elements of the local and non-local axial current for a gluon target, clarifying their connection with the axial anomaly.
In so doing, we also reconsider classic results obtained in the context of the nucleon spin sum rule as well as recent results for off-forward kinematics.
An important role is played by the infrared regulator, for which we put a special emphasis on the nonzero quark mass.
We highlight cancellations that take place between contributions from the axial anomaly and the quark mass, and we elaborate on the relation of those cancellations with the conservation of angular momentum.
\end{abstract}

\date{\today}


\end{frontmatter}

\section{Introduction}
\label{s:introduction}
Unlike the vector current, the axial current of spin-half fermions is not conserved.
In quantum chromodynamics (QCD), the non-conservation of the flavor-singlet axial current, $J_5^\mu(x) = \sum_q \bar{q}(x) \, \gamma^\mu \gamma_5 \, q(x)$, is expressed via 
\begin{align}
\partial_{\mu} J_5^{\mu}(x) = \sum_q 2 i m_q \, \bar{q}(x) \, \gamma_5 \, q(x) - \frac{\alpha_{s} \, N_f}{4 \pi} \, \textrm{Tr}\, \big( F^{\mu \nu}(x) \widetilde{F}_{\mu \nu}(x) \big) \,,
\label{e:Jmu5_divergence}
\end{align}
where $m_q$ is the quark mass, $\alpha_{\rm s}$ the strong coupling, $N_f$ the number of quark flavors, and $\widetilde{F}_{\mu\nu}$ the dual field strength tensor defined through $\widetilde{F}_{\mu \nu}(x) = \frac{1}{2} \varepsilon_{\mu \nu \rho \sigma} F^{\rho \sigma}(x)$ (with $\varepsilon^{0123} = 1$).
The second term on the r.h.s.~of Eq.~\eqref{e:Jmu5_divergence} represents the axial anomaly, which is generated through radiative corrections in quantum field theory~\cite{Adler:1969gk, Bell:1969ts, Adler:1969er, Fujikawa:1979ay}.
When considering the divergence of the axial current, the focus is mostly concentrated on the axial anomaly while the fermion-mass term is often neglected.
One motivation of the present work is to take the quark-mass term in Eq.~\eqref{e:Jmu5_divergence} into account, which can qualitatively change results as we discuss in more detail below.
\begin{figure}[t]
\begin{center}
\includegraphics[width = 0.5 \textwidth]{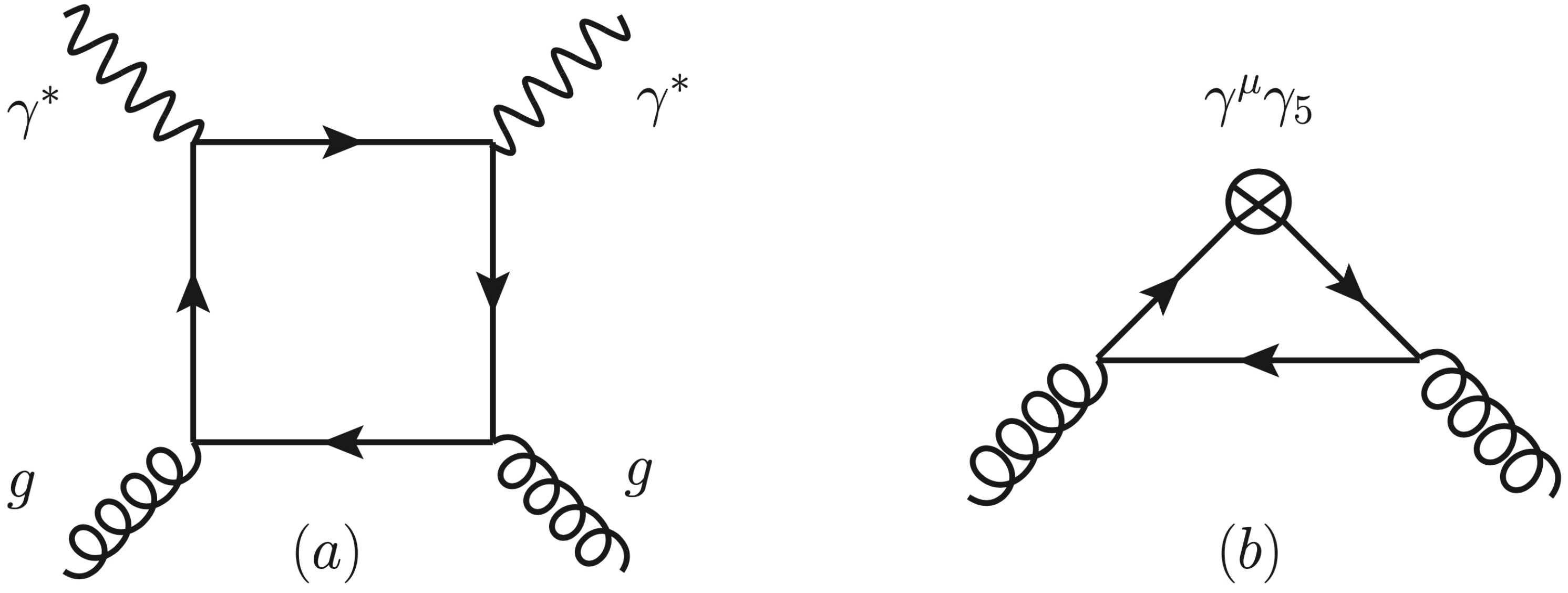}
\end{center}
\vspace{-0.4cm}
\caption{Left panel: Box diagram for the process $\gamma^{\ast} g \to \gamma^{\ast} g$. (Permutations are not shown). The imaginary part of the box diagram contributes to the DIS cross section at ${\cal O}(\alpha_{\rm s})$.
Right panel: Diagram representing the local axial current. (A second diagram, with reversed arrows on the quark lines, is not shown.)}
\label{f:box_local}
\end{figure}

It was noted early on that the axial anomaly could play an important role when trying to understand the spin structure of the nucleon in QCD~\cite{Kodaira:1979pa, Jaffe:1987sx}; see  Refs.~\cite{Anselmino:1994gn, Lampe:1998eu, Filippone:2001ux, Bass:2004xa, Aidala:2012mv} for reviews.
Research in this area intensified after the discovery of the proton `spin crisis' by the European Muon Collaboration~\cite{EuropeanMuon:1987isl}, after which works by Altarelli and Ross (AR)~\cite{Altarelli:1988nr} and by Carlitz, Collins and Mueller (CCM)~\cite{Carlitz:1988ab} received considerable attention.
These papers studied the photon-gluon fusion process, $\gamma^{\ast} g \to q \bar{q}$, in polarized inclusive deep-inelastic lepton-proton scattering (DIS), $\vec{\ell} \vec{p} \to \ell X$.
The cross section of the $\gamma^{\ast} g \to q\bar{q}$ process is represented at leading order by the imaginary part of the box diagram in Fig.~\ref{f:box_local}$\,$(a).
In the Bjorken limit $Q^2 \to \infty$ and $x_{\rm B} = Q^2 / 2 P_N \cdot q$ fixed, where $P_N \, (q)$ is the four-momentum of the nucleon (virtual photon) and $Q^2 = - q^2 > 0$, the relevant contribution of the box diagram is fully captured by the triangle diagram in Fig.~\ref{f:box_local}$\,$(b)~\cite{Carlitz:1988ab}.
It is this triangle diagram from which the axial anomaly in QCD can be computed~\cite{Adler:1969gk, Bell:1969ts, Adler:1969er}. 
Considering the full DIS process through ${\cal O}(\alpha_{\rm s})$, AR~\cite{Altarelli:1988nr} and CCM~\cite{Carlitz:1988ab} obtained 
\begin{align}
\Delta \Sigma = \Delta \widetilde{\Sigma}  - \frac{\alpha_{\rm s} \, N_f}{2 \pi} \, \Delta G \,,
\label{e:spin_anomaly}
\end{align}
in terms of the measured quark-spin contribution $\Delta \Sigma$, the `intrinsic' quark-spin contribution $\Delta \widetilde{\Sigma}$, and the amount of the proton spin due to the gluon spin $\Delta G$.
Equation~\eqref{e:spin_anomaly} was considered a potential explanation of the discrepancy between the small experimental result for $\Delta \Sigma$ and the (much) larger $\Delta \widetilde{\Sigma}$ obtained in quark models.
This development caused some excitement, especially since the $\Delta G$ term in Eq.~\eqref{e:spin_anomaly} could be attributed to the axial anomaly.

For several reasons, however, concerns came up early.
First, the pre-factor of $\Delta G$ in Eq.~\eqref{e:spin_anomaly} depends on the infrared (IR) regulator used for the calculation of the box/triangle diagram~\cite{Carlitz:1988ab, Jaffe:1989jz, Forte:1989qq, Bodwin:1989nz}.
Second, the connection of the $\alpha_{\rm s}$ term in Eq.~\eqref{e:spin_anomaly} with the anomaly in Eq.~\eqref{e:Jmu5_divergence} was questioned~\cite{Jaffe:1989jz}.
Third, the way how one factorizes the box diagram into a perturbative and a non-perturbative part can modify Eq.~\eqref{e:spin_anomaly}~\cite{Jaffe:1989jz, Bodwin:1989nz, Vogelsang:1990ug}.
(This is related to the first point about the IR regulator.)
In this paper, we mainly concentrate on the first two points.

Recently, Tarasov and Venugopalan~\cite{Tarasov:2020cwl, Tarasov:2021yll}, as well as Bhattacharya, Hatta and Vogelsang~\cite{Bhattacharya:2022xxw, Bhattacharya:2023wvy} took a fresh look at this topic.
Among other things, they pursued the idea of using a nonzero momentum transfer as an IR regulator of the box diagram with massless quarks~\cite{Jaffe:1989jz}.
They argued that, in contrast to the forward kinematics of the DIS process, off-forward kinematics would allow one to (fully) capture the physics of the anomaly. 
They furthermore argued that, in the forward limit, the anomaly would give rise to a pole term in perturbation theory, potentially endangering QCD factorization.
(Note, however, that in the latest of these papers it was shown that the anomaly contribution can very well be compatible with factorization~\cite{Bhattacharya:2023wvy}.)
The use of off-forward kinematics suggests that deeply virtual Compton scattering (DVCS) off the proton~\cite{Muller:1994ses, Ji:1996ek, Radyushkin:1996nd}, $\ell p \to \ell p \gamma$, is well suited for studying the axial anomaly.
This result was implied by Refs.~\cite{Tarasov:2020cwl, Tarasov:2021yll} and elaborated on in great detail in Refs.~\cite{Bhattacharya:2022xxw, Bhattacharya:2023wvy}.
Both the real and imaginary part of the box diagram for the process $\gamma^{\ast} g \to \gamma g$ contribute to the DVCS scattering amplitude at ${\cal O}(\alpha_{\rm s})$~\cite{Collins:1998be, Ji:1998xh}.
Here, we revisit pertinent perturbative calculations, and we confirm that for off-forward kinematics there is (also) a clear connection with the axial anomaly. 
On the other hand, we show by exploiting physical polarization vectors for the gluons that the axial anomaly does not generate a pole in such calculations when taking the forward limit.
In fact, for a nonzero quark mass there is even an exact cancellation between the anomaly and quark-mass contributions in that limit, a result which is actually required by the conservation of angular momentum.

In Sec.~\ref{s:PDF}, we calculate, to the lowest non-trivial order in perturbative QCD, the helicity-dependent parton distribution function (PDF) defined through the non-local axial current using a gluon target. 
Integrating this PDF over the momentum fraction $x$ of the quark provides the matrix element of the local axial current in the forward limit, which gives the pre-factor of $\Delta G$ in Eq.~\eqref{e:spin_anomaly}~\cite{Carlitz:1988ab}.
We discuss the dependence of the result on the IR regulator and, in particular, confirm the nonzero CCM result for off-shell gluons~\cite{Carlitz:1988ab}.
In Sec.~\ref{s:Local}, we consider the matrix element of the local axial current for arbitrary momentum transfer, and we discuss its relation with the matrix element of the anomaly and the quark-mass term in Eq.~\eqref{e:Jmu5_divergence}.
We show that for on-shell gluons, the axial current vanishes for forward kinematics.
(If the calculation is not gauge invariant the result can be nonzero, which applies to the classic AR work~\cite{Altarelli:1988nr}.)
This result can be understood as a consequence of the cancellation between the anomaly and quark-mass terms.
In Sec.~\ref{s:GPD}, we consider the helicity-dependent generalized parton distributions (GPDs)~\cite{Muller:1994ses, Ji:1996ek, Radyushkin:1996nd} that parameterize the matrix element of the non-local axial current for off-forward kinematics. 
One of the GPDs is unambiguously related to the axial anomaly, as already emphasized in Refs.~\cite{Bhattacharya:2022xxw, Bhattacharya:2023wvy}.
However, we argue that the contribution of that GPD vanishes in the forward limit when using the finite quark mass as IR regulator.
We conclude in Sec.~\ref{s:conclusions}.

\section{Parton distribution function}
\label{s:PDF}
\begin{figure}[t]
\begin{center}
\includegraphics[width = 0.25 \textwidth]{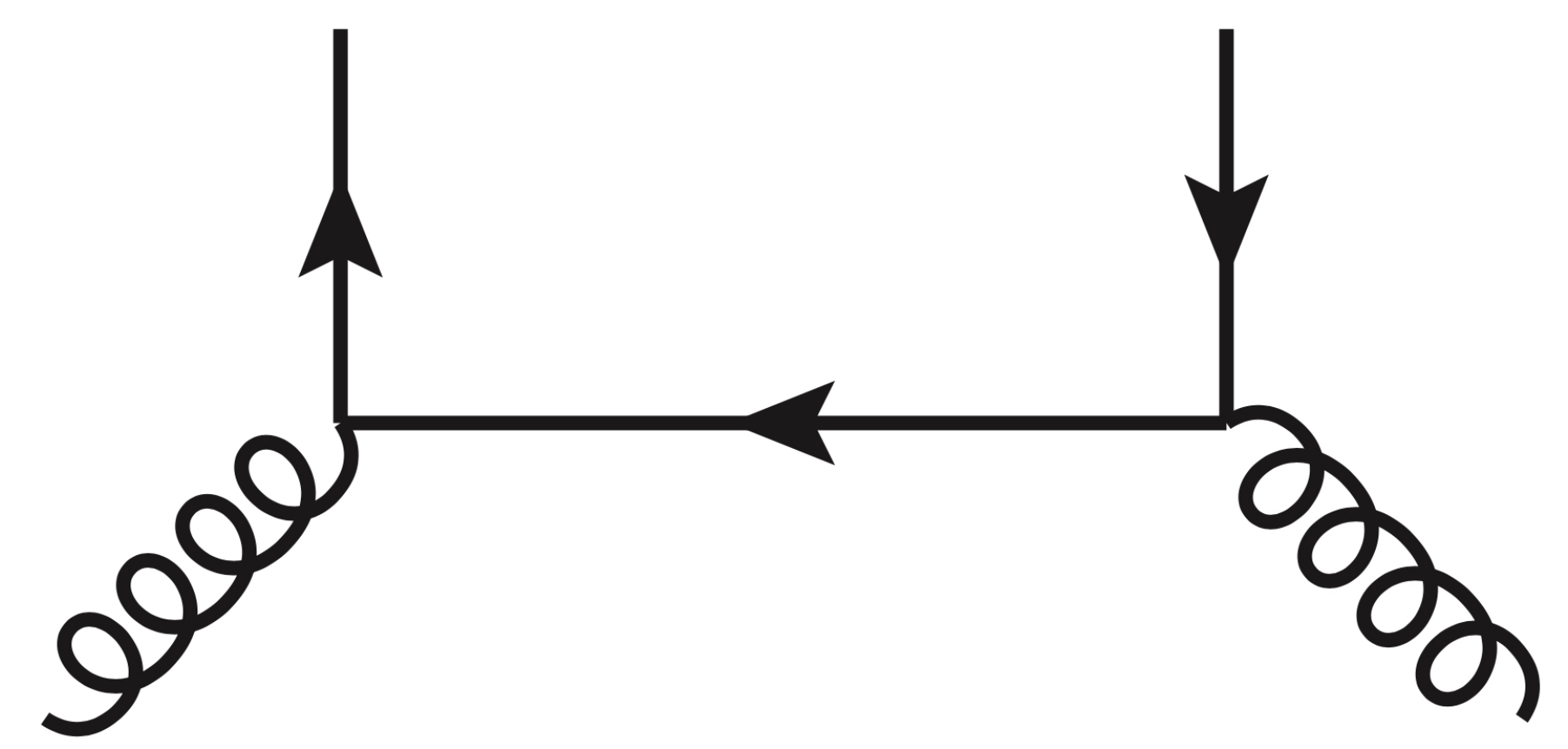}    
\end{center}
\vspace{-0.4cm}
\caption{Lowest-order diagram contributing to the PDF $g_1(x)$ defined in Eq.~\eqref{e:def_PDF_g1}.
A second diagram, with reversed arrows on the quark lines, is not shown.
The displayed diagram provides the PDF in the region $0 \le x \le 1$, while the second graph gives the result for $-1 \le x \le 0$.} 
\label{f:PDF_GPD}
\end{figure}
We consider the light-cone operator of the axial quark current, evaluated between gluon states, and we define the corresponding PDF 
(denoted by $g_1(x)$) according to\footnote{For a generic four-vector $a = (a^0, a^1, a^2, a^3)$, we define the light-cone components via $a^+ = \frac{1}{\sqrt{2}}(a^0 + a^3)$, $a^- = \frac{1}{\sqrt{2}}(a^0 - a^3)$, and $\vec{a}_\perp = (a^1, a^2)$.}
\begin{align}
\Phi_{\lambda \lambda'}^{[\gamma^+ \gamma_5]}(x) = \int \frac{dz^-}{4\pi} \, e^{i k \cdot z} \, \langle g (p, \lambda') \, | \, \bar{q}(- \tfrac{z}{2}) \, \gamma^+ \gamma_5 \, {\cal W}(- \tfrac{z}{2}, \tfrac{z}{2}) \, q(\tfrac{z}{2}) \, | \, g (p, \lambda) \rangle \big|_{z^+ = 0, \vec{z}_\perp = \vec{0}_\perp} 
= - \frac{i}{p^+} \, \varepsilon^{+ \, \polini \, \polfin p} \, g_1(x) \,,
\label{e:def_PDF_g1}
\end{align}
with $x = k^+ / p^+$.
Here $\lambda \, (\lambda')$ indicates the polarization state of the incoming (outgoing) gluon, ${\cal W}(- \tfrac{z}{2}, \tfrac{z}{2})$ represents the Wilson line rendering the bilocal quark operator gauge invariant, and $\varepsilon^{+ \, \polini \, \polfin p} = \varepsilon^{+ \, \mu \, \nu \, \rho} \, \polini_{\mu} \, \polfin_{\nu} \, p_{\rho}$.\footnote{To keep the notation simple, we do not show the dependence of the PDF on the renormalization scale, and we write $\polinimu{\mu}$ instead of $\polinimu{\mu} (\lambda)$ (and likewise for the outgoing gluon).  
Furthermore, in what follows we mostly consider one quark flavor only, and we denote the quark mass by $m$.}
It is worth emphasizing that, due to the forward kinematics, $\varepsilon^{+ \, \polini \, \polfin p}$ is the only structure that appears when parameterizing the light-cone correlator $\Phi_{\lambda \lambda'}^{[\gamma^+ \gamma_5]}(x)$.
Evaluating that structure for different polarization states of the gluons provides
\begin{align}
g_1(x) =\frac{1}{2} \left(\Phi_{++}^{[\gamma^+ \gamma_5]}(x) -
\Phi_{--}^{[\gamma^+ \gamma_5]}(x) \right) \,,
\label{e:g1_helicity}
\end{align}
where $+ \, (-)$ means positive (negative) helicity. 
This implies, in particular, that the matrix element in Eq.~\eqref{e:def_PDF_g1} is nonzero only if the helicity of the incoming and outgoing gluon is the same.
A gluon helicity flip is forbidden due to conservation of angular momentum. 
(Note that the two quark fields of the light-cone operator in Eq.~\eqref{e:def_PDF_g1} have the same helicity.  Even if their helicity was different, a gluon helicity-flip is not allowed for forward kinematics since it requires a change of two units of angular momentum.)
We will come back to this point below.

We compute $g_1$ to lowest non-trivial order in perturbative QCD (see Fig.~\ref{f:PDF_GPD}) by keeping a finite quark mass and space-like off-shellness $p^2 < 0$ for the gluon.
We use dimensional regularization (DR) to deal with ultraviolet (UV) divergences.
To define the matrix $\gamma_5$ in DR, we employ the Larin scheme~\cite{Akyeampong:1973xi,Larin:1993tq}, along with the replacement $\gamma^\mu \gamma_5 \to \frac{1}{2} (\gamma^\mu \gamma_5 - \gamma_5 \gamma^\mu)$; see, for instance, Ref.~\cite{Moch:2015usa}.
For the positive-$x$ region, we find
\begin{align}
g_1(x; m, p^2) = 
\frac{\alpha_{\rm s}}{4 \pi} \,  \Bigg[ \Bigg( \frac{1}{\varepsilon} - \ln \frac{m^2 - p^2 x (1 - x)}{\bar{\mu}^2} \Bigg) \, (2x - 1) + \frac{p^2 x (1 - x)}{m^2 - p^2 x (1 - x)} \Bigg] + {\cal O}(\varepsilon) \qquad 0 \le x \le 1 \,,
\label{e:g1_result}
\end{align}
where $\bar{\mu}^2 = 4\pi e^{- \gamma_E} \mu^2$,
with $\mu$ being the DR scale and $\gamma_E$ the Euler constant.
(The corresponding result in quantum electrodynamics is obtained by $\alpha_{\rm s} / 2 \to \alpha_{\rm em}$, where $1/2$ is the color factor of the QCD diagram.)
The result for the negative-$x$ region can be found by substituting $x \to -x$ in Eq.~\eqref{e:g1_result}.
The UV divergence of $g_1$ is reflected by the $1/\varepsilon$ pole.
Both $m$ and $p^2$ serve as IR regulators.
Notice that in order to obtain a IR-finite result just one of these two regulators is needed.
(However, when using nonzero $p^2$ only, a logarithmic endpoint singularity emerges.)

The lowest moment of $g_1$, which provides the local axial current for forward kinematics, is UV-finite and given by
\begin{align}
\int_{-1}^{1} dx \, g_1(x; m, p^2) = \frac{\alpha_{\rm s}}{2\pi} \, \Bigg[ - 1 + \int_0^1 dx \, \frac{2 m^2 (1 - x)}{m^2 - p^2 x (1 - x)} \Bigg]
= \frac{\alpha_{\rm s}}{2 \pi} \,  \Bigg[ - 1 + \frac{2}{\sqrt{\eta \, (\eta + 4)}} \, \ln \frac{\sqrt{\eta + 4} + \sqrt{\eta}}{\sqrt{\eta + 4} - \sqrt{\eta}} \Bigg] \,.
\label{e:g1_moment}
\end{align}
The expression after the first equal sign in Eq.~\eqref{e:g1_moment} matches Eq.~(12) of the CCM paper~\cite{Carlitz:1988ab}.
The moment of $g_1$ depends on $m$ and $p^2$ only through the ratio $\eta = -p^2 / m^2 > 0$.
It is IR-finite but does depend on the numerical values of the IR regulators.
For the limits $\eta \to 0$ (corresponding to $p^2 \to 0$ and finite $m$, or finite $p^2$ and $m \to \infty$) and $\eta \to \infty$ (corresponding to finite $p^2$ and $m \to 0$, or $p^2 \to \infty$ and finite $m$), we find
\begin{align}
\int_{-1}^{1} dx \, g_1(x; m, p^2) &= \frac{ \alpha_{\rm s}}{2 \pi} \, \Bigg[ - \frac{\eta}{6} + {\cal O}\big(\eta^2 \big) \, \Bigg] \,\stackrel{\eta \, \to \, 0}{\to} \, 0 \,, \qquad \qquad 
\label{e:g1_moment_p20}
\\
\int_{-1}^{1} dx \, g_1(x; m, p^2) &= \frac{ \alpha_{\rm s}}{2 \pi} \,
\Bigg[ - 1 + \frac{2}{\eta} \ln \eta + {\cal O} \bigg(\frac{1}{\eta^2} \bigg) \, \Bigg] \, \stackrel{\eta \, \to \, \infty}{\to} \, - \frac{\alpha_{\rm s}}{2 \pi} \,,
\label{e:g1_moment_m0}
\end{align}
respectively.
Multiplying the result in Eq.~\eqref{e:g1_moment_m0} by the number of quark flavors $N_f$ provides the pre-factor of $\Delta G$ in Eq.~\eqref{e:spin_anomaly}; see also Ref.~\cite{Carlitz:1988ab}.
We repeat that the dependence of the lowest moment of $g_1$ on the IR regulator gave rise to extensive discussions~\cite{Carlitz:1988ab, Jaffe:1989jz, Forte:1989qq, Bodwin:1989nz, Lampe:1998eu}, putting into question Eq.~\eqref{e:spin_anomaly} as a viable explanation of the proton spin crisis.
Here we do not elaborate further on this point.
In the next section, however, we make explicit the connection between the result in Eq.~\eqref{e:g1_moment} and the operators on the r.h.s.~of Eq.~\eqref{e:Jmu5_divergence}.

\section{Local axial current}
\label{s:Local}
We proceed to the discussion of the matrix element of the local axial current $J_5^\mu(x)$ for nonzero momentum transfer.
We first consider the matrix element of the divergence of the current,
\begin{align}
\langle g (p', \lambda') \, | \, \partial_\mu J_5^\mu (0)\, | \, g (p, \lambda) \rangle
= -2 \, \varepsilon^{\polini \, \polfin P \, \Delta} D(\Delta^2) = - 2 \, \varepsilon^{\polini \, \polfin P \, \Delta} \big( D_a(\Delta^2) + D_m(\Delta^2) \big) \,,
\label{e:me_Jmu5_divergence}
\end{align}
with $P = (p + p')/2$ and $\Delta = p' - p$.
The quantity $D_a \; (D_m$) is the contribution of the anomaly (mass) term in Eq.~\eqref{e:Jmu5_divergence}.
(The factor $-2$ on the r.h.s.~of Eq.~\eqref{e:me_Jmu5_divergence} has been introduced for later convenience.)
Because of the structure $\varepsilon^{\polini \, \polfin P \, \Delta}$ in Eq.~\eqref{e:me_Jmu5_divergence}, off-forward kinematics is required to obtain a nonzero result.
However, this does not imply that the moment of $g_1$ in Eq.~\eqref{e:g1_moment} is unrelated to the anomaly (nor to the quark mass term), as we will discuss below in this section.

We evaluate the matrix element in Eq.~\eqref{e:me_Jmu5_divergence} for two cases: (i) arbitrary $\Delta^2$, nonzero quark mass, on-shell gluons; (ii) zero $\Delta^2$, nonzero quark mass, off-shell gluons $(p^2 = p'^2 < 0)$.
The anomaly contribution follows from a tree-level calculation, while the quark-mass term is obtained by evaluating the triangle diagram for the pseudo-scalar current.
(This triangle diagram is UV finite.  Its evaluation is simpler than the direct calculation of the axial current where a UV regulator is needed at intermediate steps.)

\begin{figure}[t]
\begin{center}
\includegraphics[width = 0.50 \textwidth]{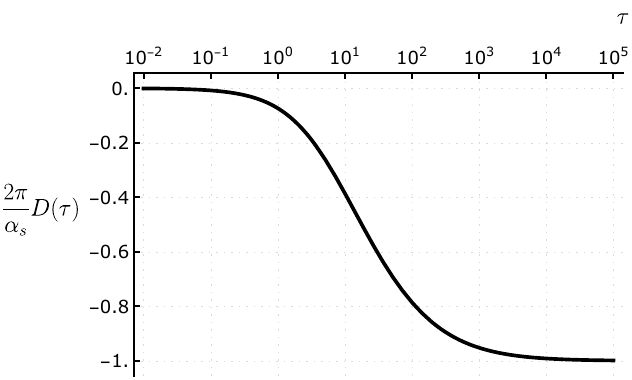}    
\end{center}
\vspace{-0.4cm}
\caption{$(2 \pi / \alpha_{\rm s}) \, D(\Delta^2, m, 0) $ as a function of $\tau$, based on the results in Eq.~\eqref{e:D_case1}.}
\label{f:D}
\end{figure}

For the case (i), we find
\begin{align}
D_a(\Delta^2; m, 0) = - \frac{\alpha_{\rm s}}{2 \pi} \,,
\qquad
D_m(\Delta^2; m, 0) = \frac{\alpha_{\rm s}}{2 \pi} \, \frac{1}{\tau} \, \ln^2 \frac{\sqrt{\tau + 4} + \sqrt{\tau}}{\sqrt{\tau + 4} - \sqrt{\tau}} \,,
\label{e:D_case1}
\end{align}
with $\tau = - \Delta^2 / m^2$.
Note that the result for $D_a$ does not depend on the momentum transfer or the IR regulator.
In the limits $\tau \to 0$ and $\tau \to \infty$, we have
\begin{align}
D(\Delta^2; m, 0) &= \frac{ \alpha_{\rm s}}{2 \pi} \, \Bigg[ - \frac{\tau}{12} + {\cal O}\big(\tau^2 \big) \, \Bigg] \,\stackrel{\tau \, \to \, 0}{\to} \, 0 \,, \qquad \qquad 
\label{e:D_Delta0}
\\
D(\Delta^2; m, 0) &= \frac{ \alpha_{\rm s}}{2 \pi} \,
\Bigg[ - 1 + \frac{1}{\tau} \ln^2 \tau + {\cal O} \bigg(\frac{1}{\tau^2} \bigg) \, \Bigg] \, \stackrel{\tau \, \to \, \infty}{\to} \, - \frac{\alpha_{\rm s}}{2 \pi} \,,
\label{e:D_m0}
\end{align}
respectively.
It is very interesting that for $\tau \to 0$, there is an exact cancellation between $D_a$ and $D_m$, a result that has been known for quite some time; see, e.g., Ref.~\cite{Adler:2004qt}.
(Generally, if the quark mass is much larger than any other scale, the matrix element in Eq.~\eqref{e:me_Jmu5_divergence} vanishes.)
We emphasize that this cancellation exists here for any finite value of the quark mass.
On the other hand, for $\tau\to \infty$ (corresponding to $\Delta^2$ finite and $m \to 0$) the divergence of the axial current is exclusively determined by the anomaly, as expected.
Given the qualitative difference between Eq.~\eqref{e:D_Delta0} and Eq.~\eqref{e:D_m0}, and the fact that in nature quarks have a finite mass, it is prudent to keep the quark-mass term in Eq.~\eqref{e:Jmu5_divergence}.
In the next section on GPDs, we will add further discussion related to this point.
The function $(2 \pi / \alpha_{\rm s})\, D(\Delta^2; m, 0)$ is displayed in Fig.~\ref{f:D}.
It deviates from the value $-1$ due to the contribution of the mass term, which is non-negligible over a significant range of $\tau$.

For the case (ii), we obtain (see also Ref.~\cite{Anselmino:1994gn})
\begin{align}
D_a(0; m, p^2) = - \frac{\alpha_{\rm s}}{2 \pi} \,,
\qquad
D_m(0; m, p^2) = \frac{\alpha_{\rm s}}{2 \pi} \, \frac{2}{\sqrt{\eta \, (\eta + 4)}} \, \ln \frac{\sqrt{\eta + 4} + \sqrt{\eta}}{\sqrt{\eta + 4} - \sqrt{\eta}} \,.
\label{e:D_case2}
\end{align}
These expressions for $D_a$ and $D_m$ agree exactly with the two terms of the result in Eq.~\eqref{e:g1_moment}, strongly suggesting that---despite the forward kinematics---the nonzero result of CCM for the lowest moment of $g_1$~\cite{Carlitz:1988ab} indeed has a robust connection with the anomaly and with the quark-mass term in Eq.~\eqref{e:Jmu5_divergence}. 
Below, we will make this connection explicit by considering the local axial current (rather than its divergence).
Before that, the $\eta \to 0$ and $\eta \to \infty$ limits of $D(0; m, p^2)$ are given by the corresponding expressions for $\int dx \, g_1$ in Eqs.~\eqref{e:g1_moment_p20},~\eqref{e:g1_moment_m0}.
In particular, according to Eq.~\eqref{e:g1_moment_p20}, for $\eta \to 0$ there is again an exact cancellation between the anomaly and the quark-mass terms in Eq.~\eqref{e:D_case2}.

Moving on to the matrix element of the local axial current itself, we first elaborate on its general structure (see also, e.g., Refs~\cite{Pasechnik:2005ae,Jegerlehner:2005fs} and references therein),
\begin{align}
\Gamma_5^\mu = \langle g(p', \lambda') \, | \, J_5^\mu(0)  \, | \, g(p, \lambda) \rangle 
= \sum_{i = 1}^3  \, G_i(\Delta^2) \, A_i^\mu
\,,
\label{e:axial_current}
\end{align}
where the vectors $A_i^\mu$, which are multiplied by the form factors $G_i$, are defined as
\begin{align}
A_1^\mu = - 2i \, \varepsilon^{\mu \, \polini \, \polfin P} \,, \qquad
A_2^\mu = \frac{2i}{\Delta^2} \, \Delta^\mu \, \varepsilon^{\polini \, \polfin P \, \Delta} 
\,, \qquad
A_3^\mu = \frac{4i}{\Delta^2} \, \big( \polini \cdot P \, \varepsilon^{\mu \, \polfin  P \, \Delta} + \polfin \cdot P \, \varepsilon^{\mu \, \polini \, P \, \Delta} \big)
\,.
\label{e:A_i}
\end{align}
(For simplicity, we do not display the dependence of $\Gamma_5^\mu$ and the $A_i^\mu$ on $\lambda$ and $\lambda'$.)
The $A_i^\mu$ are symmetric under the exchange $\polini \leftrightarrow \polfin$,~$p \leftrightarrow - p'$.
In principle, in Eq.~\eqref{e:axial_current} one could also include the vector
\begin{align}
A_4^\mu = \frac{2i}{\Delta^2} \, \big( \polini \cdot \Delta \, \varepsilon^{\mu \, \polfin P \, \Delta} - \polfin \cdot \Delta \, \varepsilon^{\mu \, \polini \, P \, \Delta} \big) \,,
\label{e:A_4}
\end{align}
but $A_4^\mu$ is not independent due to the relation
\begin{align}
A_4^\mu = - A_1^\mu + A_2^\mu \,,
\label{e:Schouten_1}
\end{align}
which follows from the Schouten identity
\begin{align}
g^{\alpha \beta} \, \varepsilon^{\mu \nu \rho \sigma}
= g^{\mu \beta} \, \varepsilon^{\alpha \nu \rho \sigma}
+ g^{\nu \beta} \, \varepsilon^{\mu \alpha \rho \sigma}
+ g^{\rho \beta} \, \varepsilon^{\mu \nu \alpha \sigma}
+ g^{\sigma \beta} \, \varepsilon^{\mu \nu \rho \alpha} \,.
\label{e:Schouten}
\end{align}
In order to proceed, we consider the Ward identity (gauge invariance) for the vector current to which the two gluons couple.
Defining $\overline{\Gamma}_5^{\, \mu} = \Gamma_5^\mu (\polini \to p)$, the Ward identity related to the initial-state gluon reads $\overline{\Gamma}_5^{\, \mu} = 0$. 
This implies
\begin{align}
G_1 (\Delta^2) = \Bigg( 1 - 4 \frac{p^2}{\Delta^2} \Bigg) \, G_3 (\Delta^2) \,,
\end{align}
allowing us to write the general form of the axial current as
\begin{align}
\Gamma_5^\mu = G_1(\Delta^2) \, \Bigg( A_1^\mu + \frac{\Delta^2}{\Delta^2 - 4 p^2} \, A_3^\mu \Bigg) + G_2(\Delta^2) \, A_2^\mu \,.
\end{align}
(The Ward identity related to the final-state gluon does not provide any additional constraint.)
At this point we use the Lorenz condition $\polini \cdot p = \polfin \cdot p' = 0$, for which $A_3^\mu = A_4^\mu$.
We can therefore eliminate $A_3^\mu$ utilizing Eq.~\eqref{e:Schouten_1}, leading to
\begin{align}
\Gamma_5^\mu \big|_{\rm real} &= \big( G_1(\Delta^2; m, 0) + G_2(\Delta^2; m, 0) \big) \, A_2^\mu = G(\Delta^2; m, 0) \, A_2^\mu \,, \phantom{\frac{1}{1}}
\label{e:axial_current_real}
\\
\Gamma_5^\mu \big|_{\rm virtual} &= - \frac{4 p^2}{\Delta^2 - 4 p^2} \, G_1(\Delta^2; m, p^2) \, A_1^\mu 
+  \Bigg( G_2(\Delta^2; m, p^2) + \frac{\Delta^2}{\Delta^2 - 4 p^2} \, G_1(\Delta^2; m, p^2) \Bigg) \, A_2^\mu \,.
\label{e:axial_current_virtual}
\end{align}
This means that for on-shell (real) gluons, there is just one independent form factor, which multiplies the vector $A_2^\mu$ and which we denote by $G = G_1 + G_2$.
On the other hand, for off-shell (virtual) gluons the parameterization of the matrix element of the local axial current contains two form factors.

Finally, we consider the (anomalous) axial Ward identity, which provides the connection with the matrix element of the divergence of the axial current in Eq.~\eqref{e:me_Jmu5_divergence},
\begin{align}
i \Delta_\mu \, \Gamma_5^\mu = \langle g (p', \lambda') \, | \, \partial_\mu J_5^\mu (0)\, | \, g (p, \lambda) \rangle \,.
\label{e:relation_current_divergence}
\end{align}
For on-shell gluons, Eq.~\eqref{e:relation_current_divergence} readily implies 
\begin{align}
G(\Delta^2; m, 0) = D(\Delta^2; m, 0) \,,
\label{e:relation_FF_div_real}
\end{align}
with $D(\Delta^2; m, 0)$ given by the sum of the anomaly term and quark-mass term in Eq.~\eqref{e:D_case1}. 
Therefore, the local axial current for on-shell gluons is completely fixed by the divergence of the current.
We repeat that computing $D(\Delta^2; m, 0)$ is simpler than $G(\Delta^2; m, 0)$ as defined through Eq.~\eqref{e:axial_current_real}.
However, we have performed both calculations and verified that our results satisfy Eq.~\eqref{e:relation_FF_div_real}.
Using Eq.~\eqref{e:relation_current_divergence} for off-shell gluons and focusing on the limit $\Delta^2 \to 0$ leads to
\begin{align}
G_1(0; m, p^2) + G_2(0; m, p^2) = G_1(0; m, p^2)= D(0; m, p^2) \,,
\label{e:relation_FF_div_virtual}
\end{align}
with $D(0; m, p^2)$ given by the sum of the two terms in Eq.~\eqref{e:D_case2}. 
We note that by explicit calculation we find $G_2(0; m, p^2) = 0 $, giving rise to the first equal sign in Eq.~\eqref{e:relation_FF_div_virtual}.
Although for off-shell gluons and arbitrary $\Delta^2$ only one linear combination of the two form factors is related to the divergence of the current, the axial Ward identity fully fixes the matrix element of the current in the forward limit.
Equation~\eqref{e:relation_FF_div_virtual} allows us to write
\begin{align}
\lim_{\Delta^2 \, \to \, 0} \Gamma_5^\mu \big|_{\rm virtual} &= G_1(0; m, p^2) \, A_1^\mu \,.
\label{e:axial_current_virtual_forward}
\end{align}
Comparing Eq.~\eqref{e:axial_current_virtual_forward} for $\mu=+$ with Eq.~\eqref{e:def_PDF_g1} integrated over $x$ provides the relation
\begin{align}
\int_{-1}^{1} dx \, g_1(x; m, p^2) = G_1(0; m, p^2) \,.
\label{e:relation_g1_FF}
\end{align}
We have now achieved three things in connection with the classic result of the CCM paper~\cite{Carlitz:1988ab} (see Eq.~\eqref{e:g1_moment}).
First, we showed that this result is unambiguously related to the matrix element of the operators on the r.h.s.~of Eq.~\eqref{e:Jmu5_divergence}. 
Second, by using Eqs.~\eqref{e:relation_FF_div_virtual} and~\eqref{e:D_case2}, we
decomposed the result into the contributions from the anomaly and from the classical quark-mass source term, finding that the vanishing result in Eq.~\eqref{e:g1_moment_p20} is caused by an exact cancellation of these two contributions.
Third, with the help of Ward identities we derived
Eqs.~\eqref{e:relation_g1_FF} and~\eqref{e:relation_FF_div_virtual},
allowing Eq.~\eqref{e:g1_moment} to be found through a simpler, UV-finite calculation
(thus also avoiding the need to choose a $\gamma_5$ scheme).

We have derived and explained the result in Eq.~\eqref{e:g1_moment_p20} by starting from the general form of the axial current for off-shell gluons in Eq.~\eqref{e:axial_current_virtual}, then taking the limit $\Delta^2 \to 0$, followed by the limit $p^2 \to 0$.
However, it is instructive to repeat the analysis by first taking $p^2 \to 0$ and afterwards $\Delta^2 \to 0$.
This means we will start from Eq.~\eqref{e:axial_current_real}.
We will also evaluate $A_1^\mu$ and $A_2^\mu$ using physical polarization vectors of the gluons.
For this analysis, we choose the symmetric reference frame in which
\begin{align}
P = \Bigg( P^+, \, \frac{\vec{\Delta}_\perp^{\, 2}}{8 (1 - \xi^2) P^+}, \, \vec{0}_\perp \Bigg) \,, \qquad
\Delta = \Bigg( -2 \xi P^+, \, \frac{\xi \vec{\Delta}_\perp^{\, 2}}{4 (1 - \xi^2) P^+}, \, \vec{\Delta}_\perp \Bigg) \,,
\label{e:sym_frame}
\end{align}
but we emphasize that our general conclusions do not depend on this choice.
The so-called skewness variable $\xi$ defines the longitudinal momentum transfer to the gluons.
In addition to the frame-independent relations $P \cdot \Delta = 0$ and
$P^2 = - \Delta^2 / 4$, we have $\Delta^2 = - \vec{\Delta}_\perp^{\, 2} / (1 - \xi^2)$.
We employ the polarization vectors specified in Ref.~\cite{Berger:2001zb} using the light-cone gauge.
In our notation, for the (on-shell) initial-state gluon they are given by
\begin{align}
\polini_{(1)}^\mu = - \frac{1}{\sqrt{N}} \, \bigg( 2 \xi P^\mu + \Delta^\mu + \Delta^2 \, \frac{n^\mu}{2 P \cdot n} \bigg) \,, \qquad
\polini_{(2)}^\mu = + \frac{1}{\sqrt{N}} \, \frac{\varepsilon^{\mu \, n \, P \, \Delta}}{P \cdot n} \,,
\label{e:pol_vectors}
\end{align}
where $N = \vec{\Delta}_\perp^{\,2}$, and $n$ is the light-cone four-vector for which $n \cdot a = a^+$ given a generic four-vector $a$.
The polarization vectors in Eq.~\eqref{e:pol_vectors} satisfy
$\polini_{(i)} \cdot \polini_{(j)} = - \delta_{ij}$ for $i,j\in\{1,2\}$,
and
$\polini_{(i)} \cdot p = 0$.
The linear combinations $\polini_{(\pm)}= \mp \, \big( \polini_{(1)}\pm i \, \polini_{(2)} \big) / \sqrt{2}$ describe states of definite (light-cone) helicity.
The polarization vectors for the final-state gluon are obtained from Eq.~\eqref{e:pol_vectors} through the replacement $\Delta \to - \Delta$ (which implies $\xi \to - \xi$) and an overall sign change for $\polini_{(1)}^\mu$ and $\polini_{(2)}^\mu$~\cite{Berger:2001zb}.
With the notation $A_{1 \, (ij)}^\mu = -2 i \, \varepsilon^{\mu \, \polini_{(i)} \, \polfin_{(j)} P}$ etc.~we find
\begin{align}
A_{1 \, (11)}^\mu &= - \frac{2i}{(1 - \xi^2) P \cdot n} \, \varepsilon^{\mu \, n \, P \, \Delta} \,,  & 
A_{1 \, (12)}^\mu &= \frac{i}{1 + \xi} \, \bigg( 2 P^\mu - \Delta^\mu + \Delta^2 \, \frac{n^\mu}{2 P \cdot n} \bigg) \,, 
\nonumber \\
A_{1 \, (21)}^\mu &= - \frac{i}{1 - \xi} \, \bigg( 2 P^\mu + \Delta^\mu + \Delta^2 \, \frac{n^\mu}{2 P \cdot n} \bigg) \,, & 
A_{1 \, (22)}^\mu &= 0 \,,
\label{e:A1_con} \\
A_{2 \, (11)}^\mu &= 0 \,,  & 
A_{2 \, (12)}^\mu &= - i \, \Delta^\mu \,, 
\nonumber \\
A_{2 \, (21)}^\mu &= - i \, \Delta^\mu \,, & 
A_{2 \, (22)}^\mu &= 0 \,.
\label{e:A2_con}
\end{align} 
For the specific case $\mu = +$, the only nonzero expressions are
\begin{align}
A_{1 \, (12)}^+ = - A_{1 \, (21)}^+ = 2 i \, P^+ \,, \qquad
A_{2 \, (12)}^+ = A_{2 \, (21)}^+ = 2 i \, \xi P^+ \,.
\label{e:Ai_con_plus}
\end{align}
Using helicity states, we can write
\begin{align}
A_{1 \, (12)}^+ - A_{1 \, (21)}^+ = i \left(
A_{1 \, (++)}^+ - A_{1 \, (--)}^+ \right) \,, \qquad
A_{2 \, (12)}^+ + A_{2 \, (21)}^+ = i\left(
A_{2 \, (+-)}^+ - A_{2 \, (-+)}^+ \right) \,.
\label{e:Ai_con_lincomb}
\end{align}
Since the structure in front of $g_1$ on the r.h.s.~of Eq.~\eqref{e:def_PDF_g1} is nothing but $A_1^+ / 2p^+$, the result in Eq.~\eqref{e:g1_helicity} is equivalent to the first relation in~\eqref{e:Ai_con_lincomb}.
While Eq.~\eqref{e:g1_helicity} means that, for forward kinematics, the axial current can only generate a helicity-conserving transition, in Eq.~\eqref{e:axial_current_real} $A_2^\mu$ appears which, according to the second relation in~\eqref{e:Ai_con_lincomb}, implies a helicity flip.
Upon a closer inspection, we see that there is no contradiction though.
Since the PDF $g_1$ is defined for $\Delta = 0$ we first consider $\Gamma_5^+$ for this specific kinematical point. 
The second relation in~\eqref{e:Ai_con_plus} then leads to
\begin{align}
\Gamma_5^+ (\xi = 0, \vec{\Delta}_\perp) \big|_{\rm real} =
\Gamma_5^+ (\xi = 0, \vec{\Delta}_\perp = \vec 0_\perp) \big|_{\rm real} = 0 \,.
\label{e:axial_current_constraint}
\end{align}
(See also Ref.~\cite{Bodwin:1989nz} for a related discussion.)
We emphasize that this result, which is fully compatible with the conservation of angular momentum, is not based on the calculation of Feynman diagrams.
Equation~\eqref{e:def_PDF_g1} can be compatible with Eq.~\eqref{e:axial_current_constraint} only if 
\begin{align}
\int_{-1}^{1} dx \, g_1(x) \big|_{\rm real} = 0 \,,
\label{e:g1_int_vanishing}
\end{align}
which must hold for any IR regulator.
Taking as an example a nonzero quark mass as IR regulator and the result in Eq.~\eqref{e:g1_moment_p20}, we see agreement with Eq.~\eqref{e:g1_int_vanishing}.
Put differently, the steps leading to Eq.~\eqref{e:g1_int_vanishing} can be considered an alternative derivation of the result in Eq.~\eqref{e:g1_moment_p20}, which was first obtained by CCM~\cite{Carlitz:1988ab}.

Overall, we find consistent results regardless of the order that the
$\Delta^2 \to 0$ and $p^2 \to 0$ limits are taken.
However, both ways of taking these limits for the local axial current provide unique insights.
We also emphasize that Eq.~\eqref{e:g1_int_vanishing} does not imply that there is no contribution from the anomaly;
based on the discussion above, it rather means that for on-shell gluons and forward kinematics, the anomaly contribution is cancelled by the quark-mass term.
Alternatively, one could say that the anomaly exactly cancels the (classical) quark-mass term---and
that this cancellation is needed for consistency with 
Eq.~\eqref{e:axial_current_constraint}, which is a consequence of angular momentum conservation.

Before moving on, we point out that the conservation of angular momentum implies the (stronger) constraint $\Gamma_5^+ (\xi, \vec{\Delta}_\perp = \vec 0_\perp) \big|_{\rm real} = 0$.
Since $A_2^+$ can be nonzero for $\xi \neq 0$, this constraint requires that the form factor $G(0)$ vanishes, which it does for $m \neq 0$; see Eqs.~\eqref{e:relation_FF_div_real} and~\eqref{e:D_Delta0}.
Keeping the quark mass is therefore necessary to ensure the conservation of angular momentum.
In the next section, we will see the same result in the context of the GPDs.

Now we briefly comment on the work by AR~\cite{Altarelli:1988nr}, in which the box diagram for polarized DIS was analyzed.
The authors considered exact forward kinematics and on-shell gluons.
However, they neglected certain quark-mass terms in the numerator, implying that their result is not gauge invariant, as was already pointed out in Ref.~\cite{Bodwin:1989nz}. 
Deriving the general structure of the axial current for on-shell gluons without using the gauge-invariance constraint leads to 
\begin{align}
\Gamma_5^{\mu \, {\rm (ngi)}} \big|_{\rm real}
= G_1^{\rm (ngi)} \, A_1^\mu + G_2^{\rm (ngi)} \, A_2^\mu \,,
\label{e:axial_current_real_ngi}
\end{align}
that is, there are two independent form factors instead of just one as in Eq.~\eqref{e:axial_current_real}.
When taking the forward limit of  Eq.~\eqref{e:axial_current_real_ngi} we can get a nonzero result related to $G_1^{\rm (ngi)}$, while $G_2^{\rm (ngi)}$ drops out since $A_2^\mu$ vanishes.
This finding is consistent with the nonzero result obtained by AR~\cite{Altarelli:1988nr}.
(In a closely related study we computed $\int dx \, g_1$ by neglecting all quark-mass terms in the numerator, finding a nonzero result as well.)
Since Eq.~\eqref{e:axial_current_real_ngi} is in conflict with Eq.~\eqref{e:axial_current_constraint}, we consider the lack of gauge invariance a deficiency of the AR paper~\cite{Altarelli:1988nr},
although this does not put into question the importance of this pioneering work.

The last point we want to address in this section is the behavior of the vector $A_2^\mu$ when taking the forward limit.
(Our discussion here also applies to $A_3^\mu$ and $A_4^\mu$.)
Because of Eqs.~\eqref{e:axial_current_real} and~\eqref{e:relation_FF_div_real}, this vector is associated with the axial anomaly (and the quark-mass term) appearing in the divergence of the axial current. 
The factor $1/\Delta^2$ in the definition of $A_2^\mu$ gave rise to extensive discussions about an `anomaly pole' that would emerge for $\Delta^2 \to 0$; see, in particular, Refs.~\cite{Tarasov:2020cwl, Tarasov:2021yll, Bhattacharya:2022xxw, Bhattacharya:2023wvy}.
However, when evaluated for physical polarization vectors, according to~\eqref{e:A2_con} one finds either zero or a finite result, depending on the polarization state of the gluons.

\section{Generalized parton distributions}
\label{s:GPD}
In this section we return to the light-cone operator of the non-local axial current, now evaluated between gluon states with different momenta.
Here we exclusively consider the case of on-shell gluons.
Using the Schouten identity in Eq.~\eqref{e:Schouten}, the Ward identity for both gluons, as well as $\polini \cdot p = \polfin \cdot p' = 0$, we find
\begin{align}
F_{\lambda \lambda'}^{[\gamma^+ \gamma_5]}(x,\Delta) &= 
\int \frac{dz^-}{4\pi} \, e^{i k \cdot z} \, \langle g (p', \lambda') \, | \, \bar{q}(- \tfrac{z}{2}) \, \gamma^+ \gamma_5 \, {\cal W}(- \tfrac{z}{2}, \tfrac{z}{2}) \, q(\tfrac{z}{2}) \, | \, g (p, \lambda) \rangle \big|_{z^+ = 0, \vec{z}_\perp = \vec{0}_\perp}
\nonumber \\
&= \big( B_1 - B_2 + \xi B_3 + B_4 \big) \, H_1(x, \xi, \Delta^2) + B_2 \, H_2(x, \xi, \Delta^2) \,,
\label{e:def_GPDs}
\end{align}
with the two GPDs $H_1$ and $H_2$.
Furthermore, we have $B_i = A_i^+ / (2P^+)$ for $i = 1,2$, and 
\begin{align}
B_3 = -\frac{2i}{\Delta^2 \, P^+} \, \big( \polini \cdot P \, \varepsilon^{+ \, \polfin  P \, \Delta} - \polfin \cdot P \, \varepsilon^{+ \, \polini \, P \, \Delta} \big) 
\,, \qquad
B_4 = -\frac{i}{2(P^+)^2} \, \big( \polinimu{+} \, \varepsilon^{+ \, \polfin  P \, \Delta} + \polfinmu{+} \, \varepsilon^{+ \, \polini \, P \, \Delta} \big) \,.
\end{align}
To arrive at the decomposition in Eq.~\eqref{e:def_GPDs} we followed to some extent the classification of deuteron GPDs presented in Ref.~\cite{Berger:2001zb}.
We agree with Refs.~\cite{Bhattacharya:2022xxw, Bhattacharya:2023wvy} that the matrix element in Eq.~\eqref{e:def_GPDs} defines (just) two GPDs.
The only difference compared to those papers is that the structure in front of $H_1$ in Eq.~\eqref{e:def_GPDs} is gauge invariant.
In analogy with Eq.~\eqref{e:g1_helicity}, we can address the GPDs through specific helicity combinations,
\begin{align}
H_1(x, \xi, \Delta^2) =\frac{1}{2(1-\xi^2)} \Big(F_{++}^{[\gamma^+ \gamma_5]}(x, \Delta) - F_{--}^{[\gamma^+ \gamma_5]}(x, \Delta) \Big) 
\,, \qquad
H_2(x, \xi, \Delta^2) = - \frac{1}{2\xi} \Big(F_{+-}^{[\gamma^+ \gamma_5]}(x, \Delta) - F_{-+}^{[\gamma^+ \gamma_5]}(x, \Delta) \Big) \,.
\label{e:GPDs_helicity}
\end{align}
While $H_1$ (like $g_1$) is associated with helicity-conserving transitions, a helicity flip is needed in the case of $H_2$.
However, for $\vec{\Delta}_\perp = \vec 0_\perp$, a gluon helicity flip is forbidden due to the conservation of angular momentum (see also the discussion after Eq.~\eqref{e:g1_helicity}), and therefore the $B_2 \, H_2$ term in Eq.~\eqref{e:def_GPDs} must drop out for that kinematics.
Since $B_2 (\xi \neq 0,\vec\Delta_\perp=\vec 0_\perp) \neq 0$, the GPD $H_2$ should vanish for a vanishing transverse momentum transfer.

Before discussing our results for the GPDs, we list a few additional constraints the GPDs must satisfy.
By definition, in the forward limit the GPD correlator in Eq.~\eqref{e:def_GPDs} reduces to the one for the PDF in Eq.~\eqref{e:def_PDF_g1}, leading to
\begin{align}
H_1(x, 0, 0) = g_1(x) \,.
\label{e:rel_H1_g1}
\end{align}
Equation~\eqref{e:rel_H1_g1} assumes that the forward limit of the GPD correlator exists, which actually is not the case if a nonzero momentum transfer is taken as the sole IR regulator.
For our calculation, this situation arises if one uses $m = 0$, as we discuss in a bit more detail below.
Further constraints follow when integrating the GPD correlator in Eq.~\eqref{e:def_GPDs} over $x$, which provides the local current. 
Comparing the result with Eq.~\eqref{e:axial_current_real} we find
\begin{align}
\int_{-1}^{1} dx \, H_1(x, \xi, \Delta^2) = 0 
\,, \qquad
\int_{-1}^{1} dx \, H_2(x, \xi, \Delta^2) = G(\Delta^2) \,. 
\label{e:rel_GPD_FF}
\end{align}
The relation for $H_1$ can be considered a generalization of Eq.~\eqref{e:g1_int_vanishing}.
Furthermore, using Eq.~\eqref{e:relation_FF_div_real} we see that the GPD $H_2$ is clearly related with the axial anomaly, confirming the corresponding statement in Refs.~\cite{Bhattacharya:2022xxw, Bhattacharya:2023wvy}.

To find the two GPDs to lowest non-trivial order in perturbative QCD, we evaluate the same two Feynman diagrams that contribute to the PDF $g_1$, but now for off-forward kinematics.
The diagram in Fig.~\ref{f:PDF_GPD} contributes to the positive DGLAP region ($\xi \le x \le 1$) and the ERBL region ($-\xi \le x \le \xi$), 
whereas the second diagram contributes to the negative DGLAP region and the ERBL region.
For convenience, we introduce the variable $\kappa = \tau \, (1 - x)^2 / (1 - \xi^2)$.
Our calculation provides 
\begin{align}
H_1(x, \xi, \Delta^2; m)  &= \frac{\alpha_{\rm s}}{4 \pi}
\begin{dcases}
\frac{2x -1 - \xi^2}{1 - \xi^2}  \, \Bigg[ \frac{1}{\varepsilon} - \ln \frac{m^2}{\bar{\mu}^2} \Bigg] - 1 
\\
+ \frac{4 + (1 + \xi^2) \kappa - 2 x (\kappa + 2)}{1 - \xi^2} \frac{1}{\sqrt{\kappa (\kappa + 4)}} \ln \frac{\sqrt{\kappa + 4} + \sqrt{\kappa}}{\sqrt{\kappa + 4} - \sqrt{\kappa}}
& \phantom{-} \xi \le x \le 1 \,,  
\\[0.2cm]
- \frac{(1 - \xi)(\xi +x)}{2 \xi (1 + \xi)} \, \Bigg[ \frac{1}{\varepsilon} - \ln \frac{m^2}{\bar{\mu}^2} \Bigg] - \frac{\xi + x}{2 \xi}
\\
- \frac{\xi^2 (2 - x) - x}{2 \xi (1 - \xi^2)} \ln \Bigg[ 1 + \frac{(1 - \xi) (\xi + x) (\xi^2 + \xi (1 - x) - x) \kappa}{4 \xi^2 (1 - x)^2} \Bigg]
\\
+ \frac{4 + (1 + \xi^2) \kappa -2x (\kappa + 2)}{2 (1 - \xi^2)} \, \frac{1}{\sqrt{\kappa(\kappa + 4)}} \ln \frac{h_{+}}{h_{-}}
\; + \, (x \to -x)
& -\xi \le x \le \xi \,,
\end{dcases}
\label{e:H1_general}
\\[0.2cm]
H_2(x, \xi, \Delta^2; m)  &= \frac{\alpha_{\rm s}}{4 \pi}
\begin{dcases}
\frac{2(1 - x)}{1 - \xi^2}  \, \Bigg[ - 1 + \frac{2}{\sqrt{\kappa(\kappa + 4)}} \ln \frac{\sqrt{\kappa + 4} + \sqrt{\kappa}}{\sqrt{\kappa + 4} - \sqrt{\kappa}} \, \Bigg] 
& \phantom{-} \xi \le x \le 1 \,,  
\\[0.2cm]
\frac{2}{1 + \xi} \, \Bigg[ - \frac{\xi + x}{2 \xi} + \frac{1 - x}{1 - \xi} \frac{1}{\sqrt{\kappa(\kappa + 4)}} \ln \frac{h_{+}}{h_{-}} \Bigg]
\; + \, (x \to -x)
& -\xi \le x \le \xi \,.
\end{dcases}
\label{e:H2_general}
\end{align}
with the auxiliary functions
\begin{align}
h_\pm = 4 \xi (1-x) \pm (1-\xi) (\xi + x) \, \sqrt{\kappa} \, \big( \! \sqrt{\kappa + 4} \pm \sqrt{\kappa} \, \big) \,.
\end{align}
Results for $-1 \leq x \leq -\xi$ can be obtained from the formulas for the positive DGLAP region by substituting $x \to -x$.
The results for $H_1$ and $H_2$ are continuous at $x = \pm \xi$ (as they should be) and satisfy the constraints in Eqs.~\eqref{e:rel_H1_g1} and~\eqref{e:rel_GPD_FF}.
(We verified $H_1(x, 0, 0; m) = g_1(x; m, 0)$, since for our perturbative calculation Eq.~\eqref{e:rel_H1_g1} is meaningful only for $m \neq 0$.)
Also note that only $H_1$ is UV-divergent.

As a next step, we expand the results in Eqs.~\eqref{e:H1_general} and~\eqref{e:H2_general} for $\tau \to 0$ (corresponding to $\vec{\Delta}_\perp \to \vec 0_\perp$ and $m$ finite) and for $\tau \to \infty$ (corresponding to $\vec{\Delta}_\perp$ finite and $m \to 0$).
In both cases we keep the full dependence on $\xi$.
For $\tau \to 0$, the Taylor expansions of the GPDs read
\begin{align}
H_1(x, \xi, \Delta^2; m)  &= \frac{\alpha_{\rm s}}{4 \pi}
\begin{dcases}
\frac{2x - 1 - \xi^2}{1 - \xi^2} \Bigg[ \frac{1}{\varepsilon} - \ln \frac{m^2}{\bar{\mu}^2} -1 \Bigg] + {\cal O} \big( \tau \big) 
& \phantom{-} \xi \le x \le 1 \,,  
\\[0.2cm]
- \frac{1 - \xi}{1 + \xi} \Bigg[ \frac{1}{\varepsilon} - \ln \frac{m^2}{\bar{\mu}^2} -1 \Bigg] + {\cal O} \big( \tau \big)
& -\xi \le x \le \xi \,,
\end{dcases}
\label{e:H1_tau_0}
\\[0.2cm]
H_2(x, \xi, \Delta^2; m)  &= \frac{\alpha_{\rm s}}{4 \pi}
\begin{dcases}
- \frac{(1 - x)^3}{3 (1 - \xi^2)^2}  \, \tau + {\cal O} \big( \tau^{2} \big) 
\,\stackrel{\tau \, \to \, 0}{\to} \, 0 
& \phantom{-} \xi \le x \le 1 \,,  
\\[0.2cm]
- \frac{(\xi + x)^2 \, (\xi^2 + 2 \xi (1 - x) -x)}{12 \xi^3 (1 + \xi)^2} \, \tau + (x \to - x) + {\cal O} \big( \tau^{2} \big)
\,\stackrel{\tau \, \to \, 0}{\to} \, 0
& -\xi \le x \le \xi \,.
\end{dcases}
\label{e:H2_tau_0}
\end{align}
We highlight that $H_2$ does vanish for $\tau \to 0$, and we repeat that this result must hold, for any $x$ and $\xi$, due to the conservation of angular momentum; see the discussion after Eq.~\eqref{e:GPDs_helicity}. 
(Based on the relation for $H_2$ in Eq.~\eqref{e:rel_GPD_FF} and our result for the form factor $G(\Delta^2; m, 0)$, it was clear before explicitly computing $H_2$ that the integral of this GPD must vanish for $\tau \to 0$.
However, this does not imply that $H_2$ has to vanish for arbitrary $x$ and $\xi$.)
We proceed to the limit $\tau \to \infty$ for which we find
\begin{align}
H_1(x, \xi, \Delta^2; m)  &= \frac{\alpha_{\rm s}}{4 \pi}
\begin{dcases}
\frac{2x - 1 - \xi^2}{1 - \xi^2} \, \Bigg[ \frac{1}{\varepsilon} - \ln \Big( - \frac{\Delta^2}{\bar{\mu}^2} \Big) - \ln \frac{(1 - x)^2}{1 - \xi^2} \Bigg] - 1 + {\cal O} \Big( \frac{1}{\tau} \Big) 
& \phantom{-} \xi \le x \le 1 \,,  
\\[0.2cm]
- \frac{1 - \xi}{1 + \xi} \, \Bigg[ \frac{1}{\varepsilon} - \ln \Big( - \frac{\Delta^2}{\bar{\mu}^2} \Big) \Bigg] 
- \frac{1}{1 - \xi^2} \, \Bigg[ 2 \xi \ln (\xi^2 - x^2)   & \\
\hspace{0.0cm} + (1 + \xi^2) \ln \frac{(1 + \xi)^2}{1 - x^2} + 2 x \ln \frac{(1 - x) (x + \xi)}{(1 + x) (\xi - x)} - 4 \xi \ln (2 \xi) \Bigg] - 1 + {\cal O} \Big( \frac{1}{\tau} \Big)
&-\xi \le x \le \xi \,,
\end{dcases}
\label{e:H1_tau_infty}
\\[0.2cm]
H_2(x, \xi, \Delta^2; m)  &= \frac{\alpha_{\rm s}}{4 \pi}
\begin{dcases}
- \frac{2 (1 - x)}{1 - \xi^2}  + {\cal O} \Big( \frac{\ln \tau}{\tau} \Big) 
\,\stackrel{\tau \, \to \, \infty}{\to} \, 
- \frac{2 (1 - x)}{1 - \xi^2}
& \phantom{-} \xi \le x \le 1 \,,  
\\[0.2cm]
- \frac{2}{1 + \xi}  + {\cal O} \Big( \frac{\ln \tau}{\tau} \Big)
\,\stackrel{\tau \, \to \, \infty}{\to} \, 
- \frac{2}{1 + \xi}
&-\xi \le x \le \xi \,.
\end{dcases}
\label{e:H2_tau_infty}
\end{align}
These expanded results fully agree with those presented in Ref.~\cite{Bhattacharya:2023wvy}, where the quark mass was neglected right form the start of the calculation.
Since $m = 0$ for the leading terms in Eq.~\eqref{e:H1_tau_infty}, a (logarithmic) singularity arises in $H_1$ for $\vec{\Delta}_\perp \to \vec 0_\perp$.
This is just a manifestation of the fact that a nonzero transverse momentum transfer acts as the IR regulator.
The expressions for $H_2$ in Eq.~\eqref{e:H2_tau_infty} for both the DGLAP and ERBL regions do not diverge for $\vec{\Delta}_\perp \to \vec 0_\perp$.
However, as we already pointed out above, using those results for $\vec{\Delta}_\perp = \vec 0_\perp$ would contradict the conservation of angular momentum.
Using $m = 0$ means that $H_2$ is just given by the anomaly.
(The $x$-integral of $H_2$ in Eq.~\eqref{e:H2_tau_infty} is nothing but $D_a$ in Eq.~\eqref{e:D_case1}.)
But we have already seen above that the combination of the anomaly and the quark-mass term is needed to get consistent results for the forward limit of the local current.
It is therefore not too surprising that an issue arises for the non-local axial current when neglecting the quark mass.

Overall, we believe the most important result of this section was that---in perturbation theory---the (anomaly-related) helicity-flip contribution to the off-forward matrix element of the non-local axial current vanishes in the limit $\vec{\Delta}_\perp \to \vec 0_\perp$ when taking the quark mass into account.
We repeat that this result is required by the conservation of angular momentum. 
This result is at odds with statements in the literature suggesting an anomaly-related pole in perturbation theory~\cite{Tarasov:2020cwl, Tarasov:2021yll, Bhattacharya:2022xxw, Bhattacharya:2023wvy}.
Finally, we argue that Eq.~\eqref{e:H2_tau_0} may be considered the non-local generalization of Eq.~\eqref{e:D_Delta0}, showing that the contribution due to the quark-mass term cancels the contribution from the axial anomaly (or vice versa).

\section{Conclusions}
\label{s:conclusions}
In this work, we presented several perturbative-QCD results related to the local and non-local axial current evaluated for gluon states.
We confirmed the classic CCM result for the local current for forward kinematics for both on-shell and off-shell gluons~\cite{Carlitz:1988ab}, found a simpler way of obtaining this result, and made clear the connection of this result with the matrix elements of the axial anomaly and the quark-mass term that appear in the divergence of the axial current.
We emphasize that the axial anomaly does contribute to the axial current for forward kinematics for on-shell and off-shell gluons. 

We agree with Refs.~\cite{Jaffe:1989jz, Tarasov:2020cwl, Tarasov:2021yll, Bhattacharya:2022xxw, Bhattacharya:2023wvy} that the anomaly does play an important role in off-forward kinematics as well.
However, in the case of on-shell gluons, the anomaly contribution is associated with a gluon helicity flip which, for a vanishing transverse momentum transfer to the gluons, is forbidden by the conservation of angular momentum.
Using physical polarization vectors for the gluons and a nonzero quark mass as the IR regulator, we found an exact cancellation between the anomaly term and the quark-mass term in this kinematic limit.
We conclude that perturbative calculations related the axial current are guaranteed to be physically meaningful only if the quark mass is not neglected; see also, e.g., Ref.~\cite{Dolgov:1971ri}.
In this context, one should keep in mind that in nature quarks have finite masses. 

Recent works had argued that---in perturbative calculations---the axial anomaly would give rise to a pole when approaching the forward limit~\cite{Tarasov:2020cwl, Tarasov:2021yll, Bhattacharya:2022xxw, Bhattacharya:2023wvy}.
It was also suggested that for a full physical process involving the proton, a non-perturbative mechanism related to the generation of the mass of the $\eta'$ meson~\cite{tHooft:1976rip, Witten:1979vv, Veneziano:1979ec, Shore:1990zu, Shore:1991dv} could cancel that pole~\cite{Tarasov:2020cwl, Tarasov:2021yll}.
While we did not address a full physical process, we repeat that in perturbation theory the axial anomaly does not generate a pole.
Regarding that discussion, we point out that the argument about the conservation of angular momentum can also directly be applied to the full box diagram in Fig.~\ref{f:box_local}$\,$(a) for both the DIS and the DVCS processes.
It is known that in DVCS there can be a gluon helicity flip for forward kinematics, but only if there is a corresponding helicity flip for the photons~\cite{Hoodbhoy:1998vm, Belitsky:2000jk, Diehl:2001pm}.
If the photon helicity is conserved, a gluon helicity flip for the box is forbidden for forward kinematics. 
We therefore expect that a full calculation of the box diagram for this case---using a nonzero quark mass---will lead to the same vanishing result in the forward limit that we found in our GPD analysis.
(In this context, see also Ref.~\cite{Freund:1994ti} and references therein.)

\section*{Acknowledgement}
We have used the FeynCalc~\cite{Shtabovenko:2020gxv} and JaxoDraw~\cite{Binosi:2008ig} software.
We gratefully acknowledge a discussion with Markus Diehl.
The work of I.C.~was supported by the National Science Foundation under the Grant No.~PHY-2110472, and by a Summer Research Grant from Temple University.
The work of A.F.~was supported by the U.S. Department of Energy contract No.~DE-AC05-06OR23177, under which Jefferson Science Associates, LLC operates Jefferson Lab.
The work of A.M.~was supported by the National Science Foundation under the Grant No.~PHY-2110472, and by the U.S. Department of Energy, Office of Science, Office of Nuclear Physics under the Quark-Gluon Tomography (QGT) Topical Collaboration with Award DE-SC0023646.
The work of S.R.~was supported by the German Science Foundation (DFG), grant number 409651613 (Research Unit FOR 2926), subproject 430915355.


\bibliography{axial_anomaly}


\end{document}